# DECONSTRUCTING THE DICHOTOMOUS RELATION BETWEEN "IT ANALYSTS AND END-USERS": A CASE OF IMPLEMENTING STANDARD INDICATORS IN CAMEROON

Flora Asah, Department of Informatics, University of Oslo, florana@uio.no

Jens Kaasbøll, Department of Informatics, University of Oslo, jens@ifi.uio.no

**Abstract:** Différance and supplément are post-structuralist concepts for analyzing language in text and are most often associated with the work of Jacque Derrida. The findings after the implementation of standard health indicators in Cameroon show that staff at the peripheral level encounter multiple challenges, including lack of participation during the implementation process, and tension between staff at the peripheral level and IT staff at the central level, which result in non-use of the system. We use deconstruction to understand the root cause and the findings reveal that IT professionals and end-users are embedded in a relation of domination. That is, IT professionals are différance from end-users and end-users are supplément of IT professionals. Although end-users are portrayed as supplementary, they are supposed to manage the system, which is contradictory. This led to IT professionals having more privilege and authority over end-users. This dichotomous relation is a derivative of the organizational structure. The notion of portraying IT professionals in charge and having more authority over end-users is an avenue for conflict. The paper concludes that a HIS organizational structure where decision-making is centralized is a ground for conflict and a major roadblock of building local capacity and providing infrastructural support at the peripheral.

## 1. INTRODUCTION AND BACKGROUND

The need to provide comparable and accurate data to report on the Millennium Development Goals (MDGs) indicators compelled Cameroon to adopt and implement strategies to harmonize the national health information system (HIS) (MINSANTE-WHO, 2018). An example is the adoption of the District Health Information Software version 2 (DHIS2)[1] and the design and implementation of standard information tools. These tools include the list of standard health indicators and data collection tools known as Monthly Reporting Activity (MRA) (MINSANTE-WHO, 2018). Standard health indicators are core elements of data analysis and the standard-setting instrument used to measure the performance of healthcare services. Standards are agreed-on health information procedures (i.e., the way data are collected). They also refer to the content of HIS (i.e., the metadata and indicators) (Jacucci et al., 2006). In this study, we refer to these tools as standard information systems (SIS). An important outcome of implementing SIS is that it should build infrastructural support and resources (i.e. human capacity, and competencies) at the peripheral level (Jacucci et al., 2006). The findings of a study conducted in Cameroon show that following the implementation of SIS, staff at the peripheral level encountered multiple challenges, including a lack of skills to analyze data and non-involvement, rendering them incapable of using the system (Asah et al., 2021). Non-involvement during participation and a lack of adequate skills to use SIS could lead to resistance and non-compliance among IS users (Dwivedi et al., 2015). In IS literature, researchers and practitioners have identified these factors as major problem shaping the use of IS (Braa et al., 2002). The implementation of IS requires the active involvement and participation of all stakeholders, particularly end-users whose involvement and participation is critical to successful implementation (Grudin, 1991; Berg et al., 2003). However, participation of end-users in the IS implementation projects appears to have been less than successful (Beirne and Ramsey, 1988), due to conflicting interests, tensions, the distribution of authority (Newman and Robey, 1992), and power play (Markus, 1983; Dwivedi et al., 2015), among IS staff particularly in centralized organizational structures (Markus, 1983). Studies show that the relationship among staff during the implementation of IS leaves much to be desired (Beath, 1991). Discussions within organization on the distribution of authority mainly center on centralization and decentralization of resources as IS practitioners tend to assume that, as technical experts, they have the more power to control IS resources (Beath, 1991). While there is a large body of literature on conflicts during IS implementation of IS, little of it has addressed them from the perspective of the end-users at the peripheral level (Lin et al., 2018; Warne, 1998). In this article, we examine the relationship among IS staff during the implementation of standard indicators. Focusing on end-users, the study seeks to explore the power relation between IT professionals and end-users, and bring to the fore the meaning of their experiences. IS are socio-technical artifacts designed for local adaptation, therefore, understanding the root causes of the tension is beneficial as this could enable these issues to be addressed (Fiorelli et al., 1993), hence improving implementation and hence, the use of IS, which is one of the goals of ICD4.

---

[1] The DHIS2 is an electronic platform that facilitates data analysis and comparison. It is a web-based system that users at all levels can access provided they have Internet access and access to the platform.

The empirical setting for the study is the implementation of standard health indicators in Cameroon's health sector. The IS implementation process is one of the times that both IS groups come together and is thus an appropriate time to examine their relationship. We draw on the concepts of supplément and différance to under the relation.

## 2.    DEFINITION OF CONCEPTS AND CONCEPTUAL BACKGROUND

**2.1.    The term 'end-user'** has been used differently but in this case; it refers to potential users or anyone in the organization who will directly or indirectly use the system. This is based on the notion that the end goal of developing a system is that it is going to be useful to the consumers. An example in this case is the HIS and the end-users are staff at the peripheral level who generate, collate, analyze, and capture routine data on the IS. **IT professionals** are the designers, developers, and implementers design the IS. In this case, they are staff of the CIS[2] who also have the decision-making powers.

**2.2.    Implementation of IS** has been a topic of discussion amongst researchers in the fields of IS. Two schools of thought exist. The first views IS implementation from a pure technological stance. The main concern of this school of thought is how to offer technical solutions to integrate the systems of different backgrounds to cater to technological needs (Bostrom and Heinen, 1977). This conception is, however, criticized by the other school of thought that holds that social and organizational aspects play an equally important role in IS development. This group conceptualizes IS implementation as a socio-technical process, highlighting the need to address organizational and social issues during the development and post-development of technical issues (Bostrom and Heinen, 1977). The socio-technical approach offers a broader understanding of HIS implementation that implies that the process of implementing IS involves incorporating the technical tools and associated procedures into existing organizational routines.

**2.3.    Concepts différance and supplément**. When computers were first introduced for common use within the organization, there was an expectation shared among many observers that they would tend to centralize organizational power (Pichault, 1995). Thus, information was equated with power and the information processing capacity of computers was seen as an extension of managerial control. The dissemination of computers in the organization was accompanied by ideas that engendered a reconceptualization of managerial and organizational process, explicitly emphasizing the importance of control over IS (Bloomfield et al., 1992). This thinking has often been associated with technocratic ideas and further strengthened the feeling that computers and IS would pave the way for enhanced centralization in organization activities (Bloomfield et al., 1992). This power relation is a post-structural approach (Chiasson et al., 2012), and we use the concepts of d*ifférance* and s*upplément* explain this in our context.

---
[2] Know by the French acronym, Cellule d'Information Sanitaire

D*ifférance* is a term coined by Derrida to unveil the fundamental conceptual oppositions inherent in Western philosophy (Chiasson et al., 2012). D*ifférance* means both difference and an act of deferring. It is used to characterize how linguistic meanings are created rather than given (Norris, 1991). According to Derrida, concept(s) are understood by comparing them to similar concepts that one is familiar with. For example, one is able to distinguish colors such as blue and red because of how different each is relative to the other (Chiasson et al., 2012). In most cases, one focuses on conceptual distinction by contrasting a concept with the one that appears to be orthogonal to it. Examples are good/evil, masculine/feminine, and speech/writing (Cooper, 1989). Derrida argues that while the two elements are placed in such a dichotomous relationship that they may seem to be in simple opposition, d*ifférance* suggests that, in practice, one element of the dichotomy is often privileged over the other (Chiasson et al., 2012). The element appearing first often masks a relationship of superiority over the other. The first element is considered a primary member of the pair, while the other is considered as secondary, and often undervalued (Derrida, 1985). An example, in this case, is IT professionals/end-users. The IT professionals are considered superior while end-users are considered secondary. To "deconstruct" the opposition is to explore the tensions and contradictions between the hierarchical ordering assumed in the concept and other aspects of the concept's meaning, especially those that are indirect or implicit (Chiasson et al., 2012).

*Supplément* is defined as "an inessential extra added to something to complete itself" (Cooper, 1989). Going back to the explanation of d*ifférance* and how we attach more importance to words that appear first, exaggerating the differences between two words leads us to ignore their common roots and to undermine what may be inherently inseparable (Cooper, 1989). In such a case, the privileging of one side conceals the often critically important compensating role of the other. From our example of IT professionals/end-users, using this concept, the end-users are *supplément* of IT professionals. The interdependency of two opposing elements in a dichotomy can often be seen as the privileged element's reliance on the supplemental one for its meaning. To deconstruct, Derrida argues that if something is complete, there is no need to add any s*upplément*. *Supplément* is used to compensate for the inherent inadequacy or incompleteness of the privileged element (Beath et al., 1994; Chissson et al., 2012). Going back to our example, one would say that if IT professionals are complete, there is no need to have end-users as s*upplément*.

Deconstruction is an interpretive approach used to study language (Jones, 2003) and is considered a post-structural (Agger, 1991) approach. As an analytical strategy, it is a tool for "de-constructing" or taking things apart, including philosophical arguments, literary text, or understandings of lived experiences (Miller, 1976). It is used to examine cultural artifacts with "an eye sharply trained to look for contradictions" (Norris, 1991, p. 137), and contradictory ideas about something or someone. Contemporary social scientists who work to deconstruct text often draw on the work of Derrida[3] (Agger, 1991). Derrida sought to establish a set of rules that could be applied in the reading, writing, and interpretations of text to unearth their historical, cultural, and social construction (Agger, 1991). Deconstruction's strengths lie in revealing implicit meaning and unacknowledged biases that exist in groups of individual such as cultural, class, or gender differences, and can be used to understand the relation between IT staff and end-user in the IT domain in Cameroon. They have also been used to explore the gray areas between two groups and what can result from those perceived differences. Using these concepts highlights the social divide that is prevalent but not discussed in the IS domain.

---

[3] The most celebrated and principal exponent of deconstruction

## 3. RESEARCH SETTING AND METHODS

### 3.1 Research Setting

The empirical setting for this research is Cameroon, a low-income country located in Central Africa with a population of about 24 million in 2018. The country has a hierarchical organizational structure characterized by centralized decision-making, much bureaucracy, and poor communication channels (WFP, 2018).

#### 3.1.1 Cameroon's Health Information Systems

Routine data management in Cameroon has a long history but the country started developing its national HIS around 2013. Prior to 1995, routine data was collected and managed haphazardly, that is, each parallel health program had its own IS (ADF, 2000). For example, the North-West and the South-West regions had two projects, SESA[4] and OCEAC[5]. Each had its HIS, list of indicators, data collection forms, and computer (ADF, 2000). In 2013, the Ministry of Public Health (MoPH) made its first attempt to develop a national HIS. A presidential decree was signed that created a national department of IS, named the CIS. The CIS is charged with the responsibility of designing, implementing, and disseminating health information throughout the country. As noted previously, Cameroon has a hierarchical organizational structure that is characterized by centralized decision-making, much bureaucracy, and poor communication channels (WFP 2018). In 2017, the country adopted the DHIS2. Routine data is compiled weekly, monthly, or quarterly depending on the program, then aggregated, and recorded in the MRA booklet. The booklets are sent to the district office for validation by the information team and it is then captured on DHIS2 platform. Thereafter, those with access to DHIS2 can access the data. At the health centers, end-users have poor access to information because the books are out of date, there is no access to journals and the Internet, and the information available is not appropriate for the local situation. Many studies describe the IS as dysfunctional (Nkoa et al., 2009; Ngwakongnwi et al., 2014; Asah et al., 2017).

### 3.2 Methods

This paper is based on an interpretive phenomenological (IP) approach (Heidegger 2010). An IP approach was used as the objective of the phenomenological method to describe experiences rather than to test hypotheses (Larkin et al., 2006). Phenomenological studies are conducted when a researcher wishes to explore the perceptions and experiences of the participants from their point of view. This approach seeks to understand the meaning that individuals ascribe to their lived experiences and the researcher aims to interpret this meaning in the context of the research (Smith, 1996). Phenomenologists are concerned with understanding phenomenon from the perspective of the participants involved (Larkin et al., 2006; Van Manen, 2016). The phenomenon studied is the relation between IT professional and end-users. Pichault (1995) explains that, those actions or behaviors that lead to conflicts occur most often when IT professionals are not aware, thus, they go unresolved or undiscussed. Heidegger (2010) argues that phenomenology does not have as it object that which is visible and clearly defined, rather, it is those phenomenon that remain hidden somehow disguised which are of interest. Using IP approach enables us to interpret the meaning of the relation between both parties. Such an approach allows researchers to put themselves in the shoes of the end-users to understand their subjective experiences (Creswell, 2007), while describing the experiences as accurately as possible, and refraining from subscribing to any pre-given facts, but remaining true to the facts. Since a basic phenomenological assumption is that all human experience is structured and has meaning, the need to force a priori structured is eliminated (Larkin et al., 2006). A semi-structured interview method, as described by Colaizzi (1978) was selected as that most appropriate and powerful for obtaining current and retrospective data from participants. 25 interviews were conducted with staff in their chosen locations. Using interviews allow participants to narrate their accounts as they lived it while not limiting the researcher to the rigid format of structured interviews.

---

[4] Child Health in the South and Adamaoua
[5] Organization of Coordination for the Control of Endemic Diseases in Central Africa

| Staff interviewed | Code | No. | Brief description of Tasks |
|---|---|---|---|
| District information officer | DIO | 3 | DIO manages information at the district level |
| Facility information officer | FIO | 6 | FIO manages information either at a General/regional hospital |
| Manager of IHC | M-IHC | 7 | M-IHC & MDH manage data at a health facility |
| Manager of District Hospital | M-DH | 2 | |
| Monitoring and evaluation officer | M&E O | 4 | Information manager from parallel programs |
| Program Manager | PM | 3 | |

**Table 1: List of staff interviewed**

Interview schedules with broad questions were used to collect data from April to September 2017. The participants were selected on the basis that they attended the implementation workshop and are directly involved in the management of information at their respective health facilities. The questions focused on exploring the nature of their role and experiences during the implementation of SIS. This includes support they receive from the CIS staff, and their involvement in the implementation process. This enabled them to describe their personal and subjective experiences as freely as possible in their own words. Where necessary, the interviewer asked additional questions for clarification and elaboration.

Before the start of each interview, time was dedicated to building trust and rapport that helped to set the tone for the rest of the discussion. The aim of the study and the interviewee's rights were clarified and written informed consent was obtained. The interviews were held at the participant's office and each lasted for about 50 minutes, and was audiotaped, and transcribed verbatim. The researcher kept a diary to record comments and describe the context and behavior that may not be captured through audio recording.

Data analysis began during data collection. At the end of each day, the researcher organized the data collected and developed impressions and issues. The researcher relied on Braun and Clarke's (2006) thematic analysis to aid analysis. Data analysis was cyclical, with the researchers going back and forth, i.e., from the data to the analysis and from the analysis back to the data to refine the interpretations and to gain in-depth understanding of what the participants were saying to truly understand the world from their perspective. The researcher read each interview transcript separately to get an overall sense of the participants' lived experiences and to obtain a sense of immersion in the data. It should be noted that this paper is based on a large PhD research project that focuses on the challenges that staff experience following the implementation of SIS. However, this paper reports on the relationship among staff, which is part of the overall research project. For this paper, which scrutinizes the relationship between IS professionals and end-users, the researcher identified statements within the data that seemed most likely to achieve this objective. We identified decision-making around HIS and the organization of IS implementation and development of training content and training. The next step was to analyze the text by identifying sections that might reflect différance or supplément.

# 4. FINDINGS AND ANALYSIS

This section presents the findings supported by verbatim quotations as well as our interpretation. Reading the transcripts, we identified two factors that characterize the relationship between IT professionals and end-users.

## 4.1 Organization of SIS implementation workshop

The CIS unit makes all decision concerning the design and implementation of IS nationwide. In this era where information is a powerful resource and CIS, being the custodian of information-related activities making it a strategic and powerful. In terms of organization hierarchy, CIS is at the top of the organization structure; they have the power of command. Therefore, the hierarchical position and organization context bestowed on CIS staff make them more powerful. As a consequent, they do not want to engage in mutual negotiations (Pichault, 1995) with the end-users. This is a technocratic reasoning as technocrats believe that information is a powerful resource that many people value in any organization (Pichault, 1995), therefore, implementing IS could create a conundrum in respect of power and control (Warne, 1998).

The implementation of SIS took the form of training workshops with staff at the peripheral level. The workshops were held in the regional health delegations but were organized, supervised, and facilitated by CIS staff. For example, a series of workshops organized at the Littoral Delegation of Health were attended by end-users (district offices and program managers) from the Littoral, Northwest, and Southwest regions.

We take as a general conviction the importance of involving staff at the peripheral level throughout the design and implementation of IS. The World Health Organization (WHO) emphasizes that the process of implementing IS should be highly participative in order to generate consensus (WHO, 1995). However, in practice, the CIS makes all decisions. As shown below, this has created a paradox that has left the end-users in an indefensible position with little or no opportunity to participate in mutual negotiations.

## 4.2 The distinction between IT professionals and end-users - différance

**Decision-making -** Despite stated commitment to and recommendations on the value and participation of end-users, a close examination of the dynamics during implementation suggests a contrary interpretation. We found that, rather than creating an atmosphere to foster joint participation, the CIS' decisions limited end-users' involvement during implementation.

The end-users stated that they were delighted that the CIS is seeking to strengthen the IS by introducing a standard list of health indicators. However, they expressed the desire to be more involved in the process; for example, in designing tools because many of those who design the system do not know the context and how data is collected:

*"…the CIS people sit in Yaoundé and design tools to be used by someone working in Limbe [on the ground] for example, without them knowing how data is collected on the ground. … when it comes to information, the central level seems to impose tools"* (PM).

The fact that the CIS designed the tools and then invited end-users to the workshop without allowing them to be involved in the design and planning, highlights a distinction between the two categories of staff and illustrates a relationship of superiority. The IT professionals have control and authority, while the end-users are subordinates. The health care sector includes partners and stakeholders with different information needs. The national list of indicators should, to some extent, accommodate the needs of the different partners (Sapirie and Orzeszyna, 1995), hence the need for stakeholders to participate. When a single body makes the decisions, as in this case, the list of indicators is not representative or comprehensive. The program managers reported that the MRA is not comprehensive because the indicators represent the information needs of the managers at the central level. This exclusive decision-making by the CIS illustrates their supremacy in IS implementation.

Reinforcing the CIS' superiority over staff at the peripheral level was the title of the workshop, namely "END-USERS WORKSHOP ON DHIS2 IMPLEMENTATION". Another way of illustrating dominance was how the staff were grouped. For example, CIS staff, i.e., computer programmers, and data analysts are known as "IT professionals" and staff at the peripheral levels are grouped as "end-users". It is interesting that in the IS domain, staff are acknowledged as 'professionals' while there is no acknowledgement that many so-called end-users may also be professionals in their respective domains. For example, the peripheral level includes professional nurses, medical doctors, monitoring and evaluation specialists, and program managers who are all professionals in their respective domains. Consistent aggregation of end-users across the disciplinary and hierarchical levels to create a homogenous end-user category facilitates stereotyping and emphasizes différance. Furthermore, the use of the term end-user depicts them as technologically naïve, and suggests passivity as those who consume, rather than those who can manage and take control of the IS.

Reaffirming the distinction between IT professionals and end-users is the separation of activities. The DIO reported that CIS staff facilitated all sections of the workshops while the end-users were participants who sat down to be taught how to manage the system. Other qualified and experienced staff at the peripheral level could have been involved, but were not among the facilitators. This distinction validates the separation between the two groups. Similarly, it depicts end-users as lacking technical knowledge, and as inexperienced consumers who play a passive role while the IT professionals are technocrats with technical expertise to build IS.

**Training content –** The end-users reported that the training primarily focused on operational aspects of using the DHIS2 platform and did not cover areas that were relevant to most staff. FIO explained:

*"The training focused mainly on operational aspects of the new data collection tools; and how to capture data, report generation on the DHIS2 platform. They did not teach areas that were relevant to program managers such as data analysis"* (FIO).

It is obvious that the training end-users received was not what they needed. Had they participated in the planning process, they would have had an opportunity to explain their training needs. Instead, the IT professionals imposed what they considered would be appropriate for end-users, giving them no room to negotiate. Such practices illustrate the superiority tendencies of CIS staff over end-users and reiterate différance - a difference between IT professionals and end-users.

### 4.3　　　End-users are a supplement to IT professionals - supplément

The data presented above shows that the CIS staff and end-users are different. That is, the CIS staff teach and provide guidance to the end-users on how to manage health indicators at the peripheral level, portraying the end-users as lacking knowledge. Focusing on these differences set the rationale for why CIS staff make all decisions on the implementation of SIS. The end-users' dependence re-affirms the power of the CIS and legitimates the condescending attitude towards end-users, but also creates a contradiction when it comes to those responsible for managing the IS at the peripheral level. At the peripheral level, end-users such as facility information managers have to ensure that data is adequately and accurately collected and submitted on time. A manager explained:

*"…it is the responsibility of facility managers to train on health indicators, data quality assessment, data use to all staff responsible for data management activities at the health facilities, and on HIS"* (FM).

Another manager added:

*"At the facility, we [managers] are responsible for ensuring adequate data quality. We have to review data, validate, and sign off on it...."* (MHIC).

If end-users are portrayed as *supplément* to IT professionals, the tasks end-users perform are not supplemental. End-users play a very significant role that is almost equivalent to that of the IT professionals. If data is not properly managed at the peripheral level, managers and decision-makers within the healthcare system will not have accurate data. Considering end-users as *supplément* to IT professionals while, at the same time, expecting them to manage the IS, is contradictory. This is because, on the one hand, they are considered as passive and technologically naïve, while on the other, they are recognized as managers who are supposed to train staff on data management-related activities and to manage the IS at the peripheral level.

## 5. DISCUSSION

This study aimed to explore the relationship between IT professionals and end-users. Focusing on the process of implementing SIS within a centralized, hierarchical structure, we found evidence of contradictions and logical inconsistency in their relation. Despite the emphasis on the importance of end-users' involvement and joint participation, in terms of decision-making, distribution of tasks, and responsibilities during implementation, we identified ambivalence regarding the degree to which end-users can be expected to be true co-agents with IT professionals. The heavy emphasis on joint participation during the implementation of IS, is contradicted by the actual procedures recommended to engage end-users. Pichault (1995) explain that from the IT professional perspective, IS implementation creates a conundrum in respect of power and control in the organization because the political structure of the organization is affected (Warne, 1998). While from the end-users' perspective, the actions and behavior of the IT professionals are equivocal in their commitment to participate, and ambivalent, reinforced by the dichotomy set up between the groups. These contradictions were expressed by employing différance and supplément. In terms of différance, end-users are portrayed as naïve and passive, while the IT professionals are presented as more knowledgeable and professional, i.e., IT professionals are in charge and end-users are supplementary staff.

Using the concept of supplément, we showed that this privileged dichotomy is not sustainable in that the end-users are expected to be responsible for the outcomes of the IS. That is, they have to manage the system. This contradiction illustrates the deep confusion surrounding the relationship between IT professionals and end-users. It leaves both end-users and IT professionals in an untenable position, with end-users submissive while IT professionals are in charge of the implementation of SIS. We suggest that these characterizations are likely to undermine mutual interactions between IT professionals and end-users, and, in an interesting twist, disable end-users, leaving them ill-equipped when negotiating with IT professionals and even using the IS as in this case. As a result, end-users do not have the necessary skills and resources required to manage the IS. These findings may shed light on end users' recurrent lack of skills and resources to use IS at the peripheral level, which has been well-documented in the literature (See: Asah et al., 2021; Njuguna et al., 2019; Lippeveld, 2017; Nicol et al., 2017; Nutley et al., 2013).

Using deconstruction is valuable in exposing inconsistency and the contradictory actions and behaviors embedded in the organization where the two groups (IT professionals and end-users) interact on a daily basis. Such behaviors or actions are often not detected and hence tend to remain undiscussed or unexamined (Beath et al., 1994). In this case, deconstruction helped bring to light the dominance and contradictory relations that exist in the management of HIS. These actions or behaviors do not exist in a vacuum; they are derivative of the institutional context, which includes industry-wide structures of historical patterns of resource allocation, as well as the norms on how and where institutions operate.

In every organization, there is a relationship of interdependence between different professional groups. These relationships are often complex and tend to be managed by dominance and are sustained by the particular structures of power and norms that constitute the organizational structure (Giddens, 1984). This could lead to one interest group having more privilege and authority over another by being assigned greater legitimacy and resources, which are the same means through which dominance is perpetuated. Similarly, in the management of HIS, the interdependence between IT professionals and end-users depicts a relationship of dominance. The very notion of end-user portrays CIS staff as naturally in charge and having the authority to decree the participation of end-users. Therefore, the process of implementing SIS in Cameroon is one in which some staff is designated as IT professionals and others as end-users, with the IT professional having more power and control over the end-users.

## 6. CONCLUSIONS

This study examined the relations between IT professionals and end-users. In theorizing these relations, we found that IT professionals and end-users are embedded in a relationship of dominance exercised through actions and behaviors similar to that of différance and supplément. These actions are derivative of the technocratic ideas HIS and the hierarchical organizational structure of HIS in Cameroon, which tends to privilege one group, IT professionals over end-users.

From a theoretical point of view, the study feeds into the discourse around IS implementation and ICT4D by illustrating a novel way of examining relations within IS management and brings to light a new factor that could explain non-use of IS at the peripheral level. In addition, using a hermeneutic phenomenological focus and the concepts, différance and supplément to scrutinize the relations between IT professionals and end-users is also novel. These concepts are generally used to deconstruct published text. Using them to deconstruct unpublished text makes the analysis unique and innovative. The paper concludes that the technocratic thinking and centralized decision-making in HIS is a major roadblock to the successful implementation and use of IS and a barrier in building local capacity and providing infrastructural support at the peripheral level in centralized organizational settings.

This study has several limitations. First, deconstruction has often been used to analyze printed and published text. Second, we only interviewed end-users and the study focused on the two English-speaking regions (formerly known as Southern Cameroons). These regions have lower levels of socio-economic development than other regions in the country. Third, fieldwork was conducted amidst political unrest in the English-speaking regions of the country. Fourth, using the hermeneutic phenomenological approach, the findings presented gives the account of the participants' interpretations and the researcher's interpretation of social reality. Another researcher could have a different interpretation. While this is one possible approach to theorize the case in point, further research is required on alternative approaches.

## AUTHOR'S CONTRIBUTIONS

FN led the interviews, transcribed and analyzed the data, and kept a reflective diary throughout the process. FN is a Cameroon and a knowledge broker who has worked with healthcare providers at the peripheral level. The impact of an existing professional relationship between the interviewer and interviewees was taken into account during data analysis through the researcher's reflexive diary entries. Throughout the analysis, the themes identified were discussed with JK and he gave critical input to the manuscript.

**ACKNOWLEDGMENTS**
We would like to thank the DIOs, FIOs, Program managers, Data managers, and health facilities managers from the North and South West and the Littoral regions who were interviewed, especially Mss. Bridget Kah, Munge Hellen, and Mr. Lukong of Limbe Regional Hospital.

**Declaration of Conflicting Interests:** The authors declared no potential conflicts of interest

**Funding:** No financial support was received for the research, authorship, and/or publication of this article.

**Ethical Approval:** This study was approved by the Norwegian Center for Research Data (Reference #: 45883). Ethical approval was sought from each Regional Delegation of Health.